\documentclass[aps,prl]{revtex4}
\def\ee{\end{equation}}
\def\bea{\begin{eqnarray}}

\def\bra#1{\langle #1 |}
\def\ket#1{| #1\rangle}

\def\braket#1#2{\langle \, #1 \, | \, #2 \, \rangle}

\def\Tr{{\rm Tr}}

\usepackage{graphicx}
\DeclareGraphicsExtensions{.png}
\begin{document}

\title{Quantum State Readout, Collapses, Probes and Signals}
\author{ Adrian Kent}
\email{A.P.A.Kent@damtp.cam.ac.uk} 
\affiliation{Centre for Quantum Information and Foundations, DAMTP, Centre for
Mathematical Sciences,
University of Cambridge, Wilberforce Road, Cambridge, CB3 0WA, United Kingdom}
\affiliation{Perimeter Institute for Theoretical Physics, 31 Caroline Street
North, Waterloo, ON N2L 2Y5, Canada.}

\begin{abstract}
Theories involving localized collapse allow the possibility that
classical information could be obtained about quantum states by 
mechanisms other than the standard quantum dynamical maps defined
by positive operator valued measurements, but nonetheless without allowing superluminal signalling.  
We can model this by extending quantum theory to include hypothetical
devices that read out information about the local quantum state at a
given point, which is defined by considering unitary evolution
modified only by the effects of collapses in its past light
cone.   Like Popescu-Rohrlich boxes, these hypothetical devices
would have practical and scientific implications if realisable.
These include signalling through opaque media, probing the physics
of distant or opaque systems without needing a reflected signal
and giving detailed information about collapse dynamics
without requiring direct observation of the collapsing system. 
These potential applications add to the motivation for systematic searches 
for possible signatures of these nonstandard extensions of quantum theory,
and in particular for relevant gravitational effects, such as the validity
of semi-classical gravity on small scales.   
\end{abstract}
\maketitle
\section{Introduction}

Nature extracts classical information from quantum states, as this
classical sentence, apparently generated by the matter in the brain of one
organism purportedly described by a quantum state, and now 
represented in the mind of another, shows.   
The gravitational fields around us may be an independent example, 
if (for instance) gravity is not quantized but fundamentally described by a (quasi-)classical theory.  

According to standard textbook quantum theory (see
e.g. \cite{schumacher2010quantum}), 
the most general way of obtaining classical information
from quantum states is via positive operator valued measurements
(POVMs). 
These define a quantum operation via the Kraus representation
\begin{equation}\label{kraus}
E (\rho ) = \sum_k A_k \rho A_k^{\dagger } \, , 
\end{equation}
where $k$ labels the different classical outcomes and 
\begin{equation}
\sum_k A_k^{\dagger} A_k = I \, . 
\end{equation}
Quantum operations respect the no-signalling
principle.   For example, if $\rho_{LR}$ is a density matrix for a 
bipartite system with Hilbert space $ H_L \otimes H_R$ and $E_L$ is 
a quantum operation on the $L$ system, then 
\begin{equation}
\Tr_L ( E_L \otimes I ) (\rho_{LR} ) ) = \Tr_L ( \rho_{LR} ) \, , 
\end{equation}
so that the fact that a measurement is carried out on the $L$ subsystem has no measurable effect
on the $R$ subsystem.  Predictions for measurements on the $R$
subsystem do, of course, generally change if one conditions on the
measurement outcome at $L$, but for an observer at $R$ to do this
requires information to be sent from $L$ after the measurement. 
Hence measurements cannot be used to signal superluminally.     
Stinespring's theorem shows that quantum operations (\ref{kraus}) can
be derived from unitary maps on a larger Hilbert space, on which
the POVMs are represented as projective
measurements.    
Hence this treatment of measurement leaves the boundary between quantum system and classical information
ambiguous.    One line of thought on this, following Everett, is that
quantum evolution is fundamentally unitary, defined on a Hilbert
space for the universe.   On this view, the appearance of
classical information must be explained by the connection between
consciousness and the universal wave function: whether this explanation follows
naturally or requires ad hoc postulates is disputed \cite{saunders2010many}. 
Another view is that quantum wave function collapse is an objective
phenomenon governed by mathematically well-defined rules whose form
we can try to conjecture
(e.g. \cite{ghirardi1986unified,diosi1987universal,ghirardi1990markov,penrose1996gravity,cmdraft}). 
Both lines of thought have problems (see e.g. \cite{saunders2010many,
  pearle2019dynamical, cmdraft} for some discussions).    
It is also worth stressing that even if there were a consensus on a consistent and confirmable version
of many-worlds quantum field theory, we would still not have strong reasons
to believe it without a consistent and confirmed quantum gravity
theory.   

Here, we work with the alternative hypothesis: that POVMs are applied at definite
points in time defined by some objective collapse model.   These
collapses may be discrete or continuous in time; we assume they are
localized in space.  

This still leaves the question: might nature possibly extract classical
information in other ways than POVMs?   Could the right theory of gravity
(or possibly some future theory characterising the contents of
consciousness from quantum states,
in the way that integrated information theory \cite{oizumi2014phenomenology} attempts for
classical networks) involve different rules? 
The honest scientific answer is that, while there is
no evidence for this, we do not know.  
As noted above, POVMs are the most general class of measurements that
can alternatively be represented by unitary operations in a larger
Hilbert space.   But whether quantum theory is universally valid
is very much an open question in the context of gravity (and
indeed of consciousness).   
There is, in any case, a consistent way of extending quantum
mechanics, at least in the idealised semi-relativistic setting (see
e.g. \cite{kent2012quantum}) often used
in discussing relativistic quantum information, to allow rules that go
beyond POVMs. 
This is \cite{kent2005nonlinearity} to postulate hypothetical devices that give information about
quantum states, following rules that are defined to be compatible with
standard quantum theory and to ensure no superluminal signalling.
  
\begin{figure}
\includegraphics[width=4in,keepaspectratio=true]{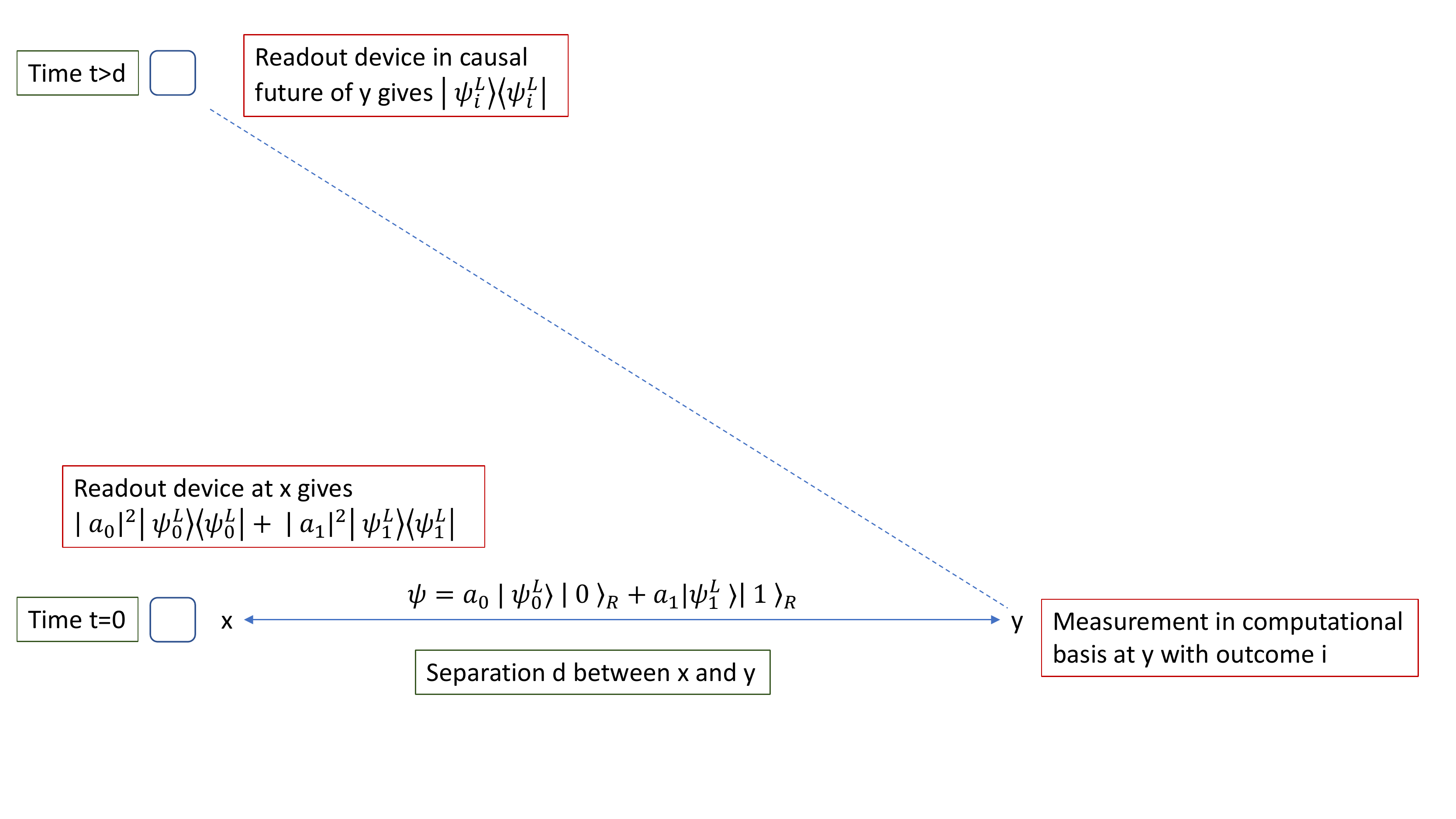}
\caption{Action of a readout device on an entangled subsystem}
\label{fig:one}
\end{figure}

A simple illustration, adequate for our discussion here, is to 
take the devices to give information about 
internal degrees of freedom such as spin, for systems of particles
with relatively negligible spread in position space, which we
treat as effectively pointlike.   
More precisely, when applied at a space-time point $x$, the devices
produce classical descriptions of (or some information about) 
the {\it local quantum state}  $\rho^{\rm loc}(x)$, 
which is defined as the local reduced density matrix at $x$ for the relevant 
degrees of freedom obtained from the quantum state on (spacelike hypersurfaces tending to) the past
light cone $\Lambda(x)$. 
We consider the devices in the context some version of quantum theory with objective localised collapses.
A device at $x$ is then sensitive to the effects of collapses
within $\Lambda(x)$, but not outside.    Any collapses arising from a measurement on a subsystem
at a point $y$ space-like separated to $x$ take place at $y$ or in its
causal future, and so do not affect the value of $\rho^{\rm loc} (x)$.
A measurement at $y$ thus has no observable effect at $x$, either via
the standard quantum state or via the devices, and so there is no
direct
way of signalling superluminally via measurements.   More generally,
it can be shown that the devices do not allow any way of 
superluminal signalling.\cite{kent2005nonlinearity,kent2005causal}.
For example, if we have two particles $L$ and $R$ at spacelike separated
points $x$ and $y$, in an entangled spin state, a spin measurement on $R$ that induces collapse at $y$ does not
affect the output of the device at $x$, but does affect the output at
points in the causal future of $y$.  (See Figure \ref{fig:one}.)

We assume here that all the readout devices 
we discuss leave the quantum state uncollapsed and unaltered.
Other possibilities are also interesting: for example, consistent
nonlinear versions of quantum theory can be defined by allowing
the local unitary evolution to depend on the readout
\cite{kent2005nonlinearity}.    
However, we focus here on the simplest case of readout devices that
have no direct effect on the quantum state.    
This is approximately the case for the simplest semi-classical gravity
models, when the gravitational self-interaction is relatively
negligible.   In these models, the semi-classical gravitational field
((\ref{gravcompavg}) below) is effectively a readout.   However, we need
to consider its precision.  

If some physical theory allowed us to actually build readout
devices, any device actually built presumably would work only to finite precision, though
it might perhaps be theoretically possible to attain arbitrarily high
precision at the cost of greater technological resources.
For example, as we discuss further below, if semi-classical gravity
were valid in some regime, we would be able to infer information 
about quantum systems in that regime from their classical
gravitational fields.    
A superposition state 
of the form $\sum_{i=0}^1 a_i \ket{x_i }$, where $\ket{x_i}$ is
a state in which a small object of mass $m$ has wave-function peaked around
$x_i$, with radius $r$ and spreads $\delta_i$ such that $r, \delta_0 ,
\delta_1 \ll | x_0 - x_1 |$, would generate
a gravitational field 
\begin{equation}\label{gravcompavg}
\Phi (y) = G m ( \frac{- | a_0 |^2 }{| x_0 - y |} + \frac{- | a_1 |^2
}{ | x_1 - y | } ) \, , 
\end{equation} 
at points $y$ with $ \min_i ( | y - x_i | ) \gg \max_i ( \delta_i )$.  
In models where measuring the gravitational field has no effect
on the quantum state, this allows estimates of the Born probabilities
$ | a_i |^2$.  In principle, by applying suitable unitaries to a qubit
and converting the computational degrees of freedom to position space,
the full form of the qubit $\sum_{i=0}^1 a_i \ket{i}$ could thus be
estimated.     

In practice, though, gravitational fields cannot be measured to infinite
precision. It might perhaps be possible to increase the precision arbitrarily by
improving the measuring devices, isolating the system as far as
possible from unknown gravitational fields and other forces, and 
so on.    Even if so, we should expect it to become greatly, perhaps
exponentially, more difficult to add to the precision beyond
a certain point.   
 
We will not generally make any detailed assumptions about this
tradeoff in our discussion, simply supposing that the relevant readout
devices can be built to sufficiently high precision.   To simplify the
notation, when the applications we consider involve 
distinguishing significantly different states, 
we treat the precision 
as effectively infinite.   

One can imagine various potentially interesting types of readout
device.   A {\it state} readout device $RD (x)$ applied at $x$ would print out a classical description
of $\rho^{\rm loc} (x)$ to given (maybe infinite) precision, expressed
in some
given basis.     An alternative idealisation, which is perhaps more elegant
(particularly for an infinite precision readout device for qubit states), is for it to represent
$\rho^{\rm loc} (x)$ physically with a classical pointer.   
An {\it expectation value} readout device $RD (A , x)$ prints out the
expectation value $\Tr ( A \rho^{\rm loc} (x) )$ of some hermitian
observable $A$ in the local state. 
A {\it stochastic eigenvalue} readout device $SRD (A, x)$ prints out 
an eigenvalue $\lambda_i$ of $A$, randomly chosen using the Born
probabilities $\Tr (P_i  \rho^{\rm loc} (x))$, where $P_i$ are the
projections onto the corresponding eigenspaces.   
A single use of a state readout device $RD (x)$ clearly allows the user to
infer the output of an expectation value readout device $RD(A,x)$ and (given a suitable
source of randomness) simulate a stochastic eigenvalue readout device $SRD(A,x)$.   
If $A$ is non-degenerate, then in principle multiple applications of either of the last
two devices, combined with appropriate unitaries, can be used
tomographically 
to produce finite precision versions of the state readout device. 

One way of viewing these readout devices is as analogous to
Popescu-Rohrlich (PR) nonlocal boxes.   
PR boxes \cite{popescu1994quantum} extend quantum theory to be (in a sense) more nonlocal,
while respecting the no-signalling principle.   
If they existed, they would allow
various information-theoretic tasks to be carried out much more
efficiently (see e.g. \cite{van2013implausible,brassard2006limit}
and Popescu's review \cite{popescu2014nonlocality}). 

Like PR boxes, readout devices extend quantum theory without violating
special relativity, and bring computational advantages.   In particular, 
both the infinite and finite precision versions of $RD$ 
would allow computational speedups.  
It follows from results of Abrams and Lloyd \cite{abrams1998nonlinear}
that 
a single infinite precision readout device that can be applied to a single qubit would allow quantum computers to
solve NP and \#P 
problems in polynomial time.    
Abrams and Lloyd's algorithm is framed in terms of an oracle that 
calculates a function $f:\{0,1\}^n \rightarrow \{0,1\}$, where the
problem is to determine whether there exists an input string $x$ such
that $f(x)=1$.  
They create a uniform superposition of all possible inputs, and
apply the oracle once, obtaining 
\begin{equation}
\psi = 2^{-n+1} \sum_{i=0}^{2^n - 1} \ket{i, f(i)} \, .
\end{equation}
With probability at least $\frac{1}{4}$, applying a Hadamard transformation to the first $n$ qubits,
followed by a measurement in the computational basis, produces the state
\begin{equation}
\psi_{n,s} = C \ket{00 \ldots 0} \otimes ( \frac{2^n - s}{2^n} \ket{0} +
\frac{s}{2^n} \ket{1} ) \, ,
\end{equation}
where $s$ is the number of values of $x$ such that $f(x)=1$ and $C$ is
the normalisation factor.     
Applying an infinite precision $RD$ to the last qubit thus gives the value of $s$
and in particular distinguishes the cases $s=0$ and $s>0$ as required. 

A sensible rough definition of a finite precision $RD$ is that, with
some given high probability $p \lesssim 1$, it is able to distinguish $\psi_{n,0}$ from
$\psi_{n,1}$ (and more generally $\psi_{n,s}$ from $\psi_{n,s'}$ for
$s \neq s'$) for $n \leq m$, with the maximal such value of $m$
effectively defining the precision.   This would efficiently solve the search problem for sets
of size up to $2^m$, a potentially valuable speed-up if $m$ is large.
It would be interesting to know if stronger results are possible.
An obstacle to straightforwardly using $RD$ within Abrams-Lloyd's 
algorithms for general $n$ is that $RD$ is defined to read out the
local density matrix, which for a single qubit readout device is
the reduced state of the relevant qubit, so that applying it 
on the penultimate qubit in 
\begin{equation}
\sum_{i=0,1} a_i \psi_i \otimes \psi_{m,i} \otimes \ket{0} 
\end{equation}
would generally produce a readout of a mixed state:
for example if $\braket{ \psi_0}{ \psi_1} = 0$ it
would produce the readout $\sum_i | a_i |^2 \ket{\psi_{m,i}} \bra{\psi_{m,i}}$. 
The readout thus cannot be directly used to 
implement transformations of the form 
\begin{equation}
\sum_{i=0,1} a_i \psi_i \otimes \psi_{m,i} \otimes \ket{0} \rightarrow \sum_{i=0,1} a_i \psi_i \otimes \psi_{m,i} \otimes \ket{i} 
\, ,    
\end{equation}
which would be desirable, since it would leave the final qubit
registering the search result on $2^m$ strings and available for
further quantum processing.    
We leave open here the question whether there may be other ways 
of using finite precision single
qubit $RD$ devices to obtain faster speedups, and the more general question
of the power of multi-qubit finite precision readout
devices, and turn to other
potential applications of readout devices. 

\section{Probes and signals}

Consider an entangled pair of subsystems at rest in state 
\begin{equation}
\ket{\Phi_+ }_{LR} = \frac{1}{\sqrt{2}} ( \ket{ 0 }_L \ket{0}_R + \ket{1}_L \ket{1}_R ) \,
,
\end{equation}
where the $L$ and $R$ subsystems are separated by distance $d$
in some inertial frame and 
$\ket{0}, \ket{1}$ are orthogonal states of each subsystem.
We take $d$ to be large compared to the wave function spread of
either subsystem and fix units with $c=1$.  
Suppose that a state readout device is applied on the $L$ subsystem.
Suppose now that at time $t=0$ (in the same frame) a projective
measurement in the $\ket{0}_R , \ket{1}_R$ basis is
applied to system $R$, which, according to the relevant collapse
hypothesis.
ensures a rapid collapse (taking time negligible compared to $d$)
onto the measured outcome state. 
The readout device at $L$ produces readout 
\begin{equation} \label{avg}
 \frac{1}{2} I_L
\end{equation}
up to time $t=d$, where $I_L$ is the uniform mixed state in
the relevant two dimensional Hilbert space.
After time $t=d$ it produces a readout 
\begin{equation} \label{outcomes}
\ket{i}_L \bra{i}_L \, ,
\end{equation}
where $i=0$ or $1$ is the outcome obtained at $R$.  
(See again Figure \ref{fig:one}.) 

An observer at L reading the readout, who knows the initial
state $\ket{\Phi_+}_{LR}$ and the locations of the subsystems, thus learns at $t=d$
that a measurement was carried out at R at time $t=0$. 
That is, the readout device gives observers at L and R, who have
previously shared the state and preagreed their locations, a means of
signalling at light speed.   The signal involves no carrier subsystem,
and so this mechanism works regardless of how opaque any intervening
material is to ordinary signals.    For example, this would allow
a simple means of signalling at light speed between antipodal points on Earth,
something which with current technology requires a strong neutrino source 
and neutrino detector.  

An expectation value readout device for the observable
$A= a_0 \ket{0}\bra{0} + a_1 \ket{1}\bra{1}$, where $a_0 \neq a_1$ 
and both are real, would work similarly: the observer at L 
would obtain readout 
$\frac{1}{2} (a_0 + a_1 ) $ up to time $t=d$ and 
then either $a_0$ or $a_1$ thereafter.  
A stochastic eigenvalue readout device for $A$, if applied repeatedly, would 
produce a random sequence of $a_0$ and $a_1$ up to time $t=d$ and
then either a sequence of $a_0$'s or a sequence of $a_1$'s.    

\subsection{Probes of distant systems}

Assume now that some specific collapse model has been empirically
confirmed, and that the readout devices function as specified with
respect to collapses in this model.   
Turning the previous observation around, if the $R$ system is sent into 
an unknown distant environment, then an observer monitoring the $L$ system
can infer, from the transition between (\ref{avg}) and
(\ref{outcomes}), that it has collapsed
at some point on the past light cone of the observed transition point.

If the $R$ system's trajectory is known (for instance, if it 
is known to travel at fixed velocity) then the location of the 
collapse can be inferred (at least to within a small region).   
In other words, properties of distant environments -- 
specifically, their propensity to cause collapse within the given
model -- can be probed, even though no particle or field perturbation
is reflected back to the observer. 

No existing technology allows this form of probing.    
Although ``interaction free'' measurement
\cite{elitzur1993quantum} or imaging (e.g. \cite{white1998interaction})
may seem somewhat analogous, they still require well-defined trajectories with non-zero
amplitudes to and from the region of the imaged object, to distinguish
between reflection and absorption (or scattering).    

\subsection{Tests of collapse hypotheses}

Suppose now that readout devices have been found to work in
combination with standard
measurements carried out by human observers using macroscopic
apparatuses.    This would be compelling evidence for some
form of objective collapse model, but the details of this
model might not immediately be fully clear.    
Measurements on the L system can then be used in order to 
obtain empirical evidence about precisely when and under what conditions
collapses occur on the R system, by exposing the R system to a variety
of potentially collapse-inducing measurement-like interactions.

For example, the R system's superposition state could be amplified
towards the macroscopic by correlating it with a variety of 
mass distribution states, in order to test and refine hypotheses
about state reduction associated with superposed mass distributions or
gravitational fields.  
Alternatively, the R system could be ``observed'' by (and so 
correlated
with the information processing states of) a variety of candidate
observers -- humans, small animals, photosynthesis mechanisms
in plants, small quantum computers, and so on -- to test and 
refine speculative hypotheses about observer- or consciousness-induced
state reduction
(e.g. \cite{cmdraft,cmtalks,kremnizer2015integrated,okon2016back}).    

Again, nothing comparable is possible with existing technology
using standard quantum theory.   Tests of collapse models currently
require either very challenging interferometry, or indirect evidence
from small violations of conservation laws or small anomalous 
wave function spreads.    These require near complete isolation
of the relevant system from environmental decoherence, which has
very similar effects.   It is also not completely clear whether violations
of conservation laws are essential corollaries of any plausible collapse
model, although they are features of all models considered to date. 

\section{The example of semi-classical gravity}

It may seem far-fetched to imagine that any of our readout devices
could be found anywhere in nature.   
However, the continuing interest in semi-classical gravity models (e.g.
\cite{moller1963theories,rosenfeld1963quantization,kibble1978relativistic,kibble1980non,carney2019tabletop,
  tilloy2016sourcing, kentgravexpt}) 
gives one reason not to dismiss the possibility.     
The literature on semi-classical gravity is inspired
by the equation
\begin{equation}\label{scg}
G_{\mu \nu} = \langle \hat{T}_{\mu \nu} \rangle  \, ,  
\end{equation}
which is easy to write but hard to interpret given (inter alia)
that the quantum matter whose stress-energy tensor appears on
the right hand side is propagating in the space-time whose metric 
determines the left hand side. 

In the non-relativistic limit with $N$ fixed particles, with
mass density operator
\begin{equation}
\hat{M}(x) = \sum_i m_i \delta (x - \hat{x_i } ) \, ,
\end{equation}
we can define (see e.g. \cite{carney2019tabletop}) the classical Newtonian potential $\Phi$ obeying 
\begin{equation}
\nabla^2 \Phi (x) = 4 \pi G \langle \hat{M}(x) \rangle \, . 
\end{equation}
Semi-classical gravity is then defined by a modified Schr\"odinger equation 
\begin{equation}\label{newtonscg}
i \frac{\partial}{\partial t} \ket{\psi} = (\hat{H}_{\rm matter} + \hat{H}_{\rm
  gravity} ) \ket{\psi} 
= (\hat{H}_{\rm matter} + \int \hat{M}(x) \Phi (x) dx ) \ket{\psi} \, .
\end{equation}
Although there are many unresolved problems with semi-classical
gravity theories \cite{kibble1978relativistic,kibble1980non,carney2019tabletop}, there
are at least ways of interpreting the non-relativistic equations \cite{kent2005nonlinearity,
  carney2019tabletop, kent2018simple, kentgravexpt} that avoid the pathological
superluminal signalling \cite{gisin1990weinberg} arising from applying
standard measurement postulates directly to (\ref{newtonscg}).   
While we do not see averaged gravitational fields from
superposed position states of large masses \cite{page1981indirect},
this leaves open the possibility that semi-classical gravity
may hold within an objective collapse model \cite{tilloy2016sourcing}
with collapses preventing macroscopically distinct mass distributions
from remaining in superposition, and with their effects on
the gravitational field propagating at light speed
\cite{kent2005nonlinearity}. 

We can replicate our earlier discussion in the context of 
semi-classical gravity by considering
an entangled pair of subsystems at rest, of the form 
\begin{equation}
\frac{1}{\sqrt{2}} ( \ket{ x }_L \ket{0}_R + \ket{x'}_L \ket{1}_R ) \,
, 
\end{equation}
where $\ket{x}, \ket{x'}$ correspond to a mesoscopic object of mass
$m$ with 
centre-of-mass well localized around the points $x, x'$ respectively,
and $\ket{0}, \ket{1}$ are orthogonal states of the other subsystem
which can be distinguished by some practical measurement method
that implies a rapid collapse onto the measured outcome state,
according to the relevant collapse hypothesis.   
We take the wave functions of the states $\ket{x}, \ket{x'}$
to have spreads $\delta, \delta' \ll | x - x' |$, and the 
L and R subsystems to be separated by $d \gg \Delta = | x - x' |$. 

Semi-classical gravity (\ref{scg}) implies that in
this state the L subsystem generates a Newtonian gravitational
potential 
\begin{equation}\label{gravavg}
\Phi(y) \approx \frac{G m}{2}  ( \frac{-1}{| x - y |} + \frac{-1}{ | x' - y | } ) 
\end{equation} 
at points $y$ with $ | y - x |, | y - x'| \gg \delta, \delta'$.   

A signal can then be sent from R to L by measuring the R subsystem, in a way that we know 
theoretically or empirically must cause a rapid
collapse, at time $0$ in the rest frame of the two subsystems.   
By hypothesis \cite{kent2005nonlinearity}, for a consistent
combination of semi-classical gravity and collapse, the Newtonian potential contributed by 
L in its vicinity remains of the form (\ref{avg}) until
approximately time $d$, when either
\begin{equation}\label{resultone}
\Phi(y) \approx \frac{-G m}{| x - y |}
\end{equation}
or 
\begin{equation}\label{resulttwo}
\Phi(y) \approx \frac{-G m}{| x' - y |} \, , 
\end{equation}
depending on the R measurement outcome.   

For suitable values of $m$ and $\Delta$, the cases (\ref{gravavg}), (\ref{resultone}) and (\ref{resulttwo}) can
be distinguished by gravitational phase interferometry \cite{kentgravexpt}, quickly
compared to $d$ (for sufficiently large $d$).  
This allows signalling, remote probing and investigation of collapse dynamics
to be carried out as above.  

\section{Searches for gravity-based readout devices}

Models involving semi-classical gravity suggest one way that readout
devices might possibly be found in nature.   More generally, any
measurable gravitational effect different from that predicted by
perturbatively quantized general relativity implies some extension
or breakdown of quantum theory, and hence the possibility of readout
devices.   Another example in this direction, which if confirmed 
would have enabled the signalling and probe mechanisms discussed
above, is the event formalism investigated by Ralph and collaborators \cite{ralph2009quantum,ralph2014entanglement,pienaar2011quantum,joshi2018space}, which predicts
anomalous gravitationally induced decoherence.  While this prediction appears
to have been falsified \cite{xu2019satellite}, it illustrates that there are
other (albeit also speculative and incomplete) theoretical motivations 
for nonlinear quantum effects connected to gravity. 

The various applications of readout devices add further motivation to
proposed tests of gravitationally-induced entanglement 
\cite{bose2017spin,marletto2017gravitationally,marshman2020locality} 
or other tests distinguishing perturbatively quantized general relativity
from possible classical models of gravity (e.g. \cite{kentgravexpt}) 
that might confirm or exclude such effects.   They also motivate
more systematic theoretical and experimental investigations.   

\begin{figure}
\includegraphics[width=4in,keepaspectratio=true]{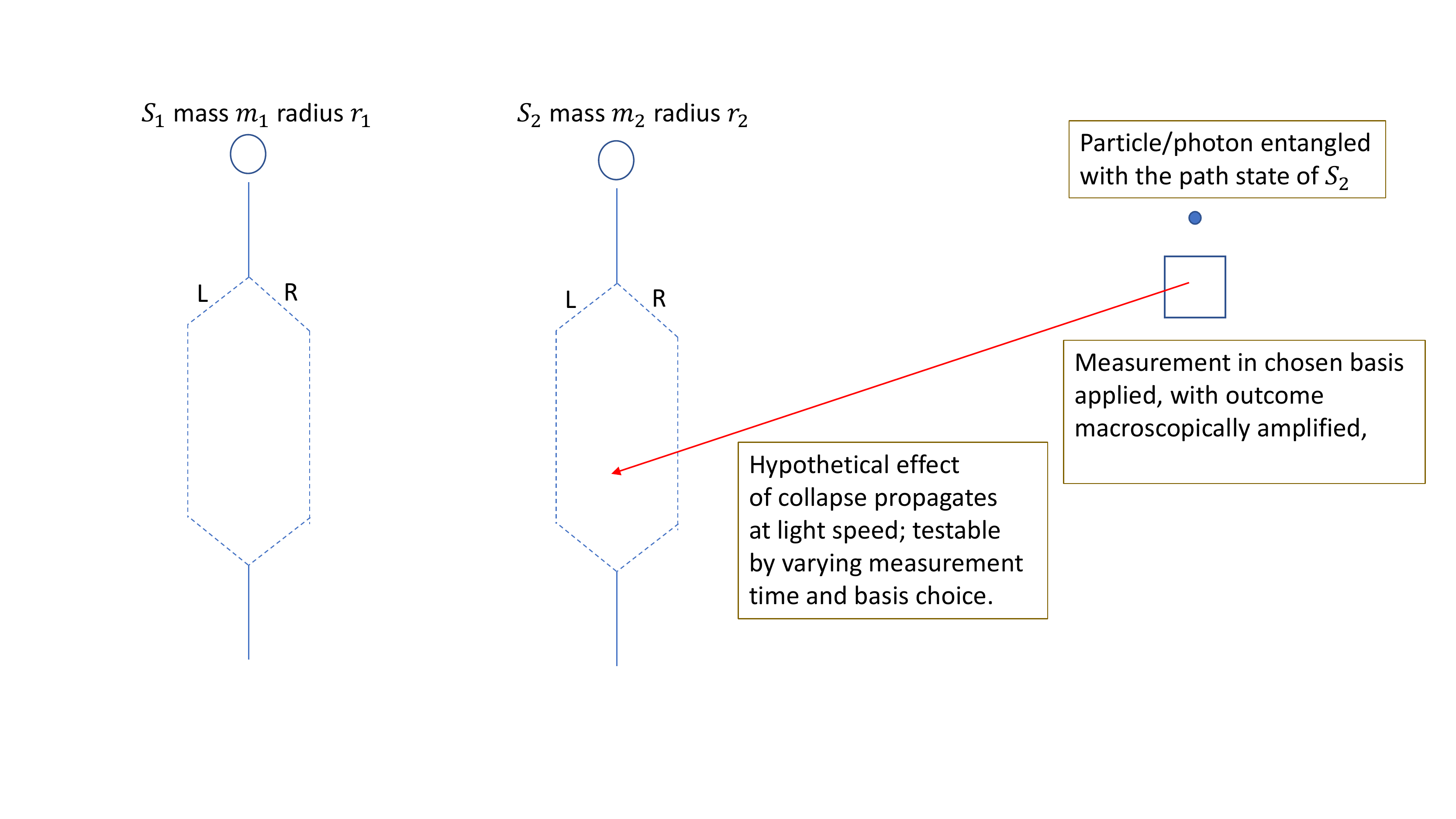}
\caption{Schematic description of interferometry experiment with
distant entanglement and measurement (not to scale).  }
\label{fig:two}
\end{figure}

A particular feature of readout device based models is that a local
system in the neighbourhood of a space-time point $x$ 
behaves differently depending on whether or not it was
entangled with a distant system up to the boundary of its past  
light cone $\Lambda_x$.   The differences could, as in the
case of semi-classical gravity, involve non-quantum effects whose 
signature might anyway be detected in experimental tests not involving distant entanglement.  
But they could also be subtle.   It would be worth systematically
searching for such effects by extending experiments such as
those of
\cite{bose2017spin,marletto2017gravitationally,carney2019tabletop,kentgravexpt}
so that at least one of the local superpositions of distinct mass
distribution is entangled with a distant system.
One can then test whether the timing of measurements (amplified
quickly to produce collapse within any plausible model) on the distant
system (within or outside the past light cone of the original
experiment) has any measurable effect.      
For example, an experiment involving two Mach-Zehnder interferometers 
producing path superpositions of masses $m_1$ and $m_2$ could be
augmented by arranging that $m_2$ is entangled with a distant
spin $\frac{1}{2}$ particle $P$ in the state
\begin{equation}
\frac{1}{\sqrt{2}} ( \ket{L} \ket{\uparrow} + \ket{R} \ket{\downarrow}
) \, . 
\end{equation}
(Equivalently, $P$ could be a photon, with its polarization degrees
of freedom entangled.)
A suitably macroscopically amplified measurement (which, by hypothesis, in the
relevant collapse model, is swiftly
collapse-inducing) on $P$ in the basis $\frac{1}{\sqrt{2}}
( \ket{\uparrow}
  \pm \ket{\downarrow} )$ would effectively place $S_2$ in one of the
states $\frac{1}{\sqrt{2}} ( \ket{L} \pm \ket{R})$.  By hypothesis,
this would occur with a time lapse (as measured in the lab frame) due to light speed transmission of the effect
of the collapse.  So, by appropriate choices of the measurement time,
it could be made to occur before, during, or after the interferometry
phase of the experiment. 
Similarly, a measurement on $P$ in the basis $\{ \ket{\uparrow},
\ket{\downarrow} \}$ would effectively place $S_2$ in one of the
states $\{ \ket{L}, \ket{R} \}$; again, by hypothesis, this 
could be made to occur before, during or after the interferometry. 
Each basis choice and each outcome defines an ensemble of experiments,
allowing a test of whether the results within that ensemble depend
in any way on the time of the distant measurement and collapse.  
(See Figure \ref{fig:two}.) 

\section{Discussion}

Nonlinear quantum effects and objective collapses are 
both speculative hypotheses, and 
it is unclear whether they can be extended to
fully consistent relativistic theories.  
Nonetheless, there are theoretical motivations
to consider both, and they have been extensively studied
(separately and together) in work on 
unifications of quantum theory and gravity.
The potential information-theoretic, technological and 
scientific applications of combining these hypotheses
has however not been fully appreciated.
For example, even the fact that semi-classical gravity would transform
our understanding of the physical basis of computing seems
not to have been noted, although it could be argued that 
this application alone should motivate far more theoretical
and experimental attention on this topic.      

The possibilities we have discussed here -- signalling, remote
probing, and alternative ways of uncovering the details of collapse mechanisms --  
add to the potential technological and scientific payoffs of 
finding any measurable nonlinear effect of the type we consider.   
From a more foundational perspective, readout device models
add to the taxonomy of possible extensions of quantum theory and of possible principles
that might identify the true theory of nature.
We hope that these observations may stimulate
further theoretical and experimental work.

\vskip10pt
\begin{acknowledgments}
This work was partially 
supported by an FQXi grant and by 
Perimeter Institute for Theoretical Physics. Research at Perimeter
Institute is supported by the Government of Canada through Industry
Canada and by the Province of Ontario through the Ministry of
Research and Innovation.   
\end{acknowledgments}

\bibliographystyle{unsrtnat}
\bibliography{collapselocexpt}{}

\begin{thebibliography}{40}
\providecommand{\natexlab}[1]{#1}
\providecommand{\url}[1]{\texttt{#1}}
\expandafter\ifx\csname urlstyle\endcsname\relax
  \providecommand{\doi}[1]{doi: #1}\else
  \providecommand{\doi}{doi: \begingroup \urlstyle{rm}\Url}\fi

\bibitem[Schumacher and Westmoreland(2010)]{schumacher2010quantum}
Benjamin Schumacher and Michael Westmoreland.
\newblock \emph{Quantum processes systems, and information}.
\newblock Cambridge University Press, 2010.

\bibitem[Saunders et~al.(2010)Saunders, Barrett, Kent, and
  Wallace]{saunders2010many}
Simon Saunders, Jonathan Barrett, Adrian Kent, and David Wallace.
\newblock \emph{Many {W}orlds?: {E}verett, Quantum Theory, \& Reality}.
\newblock Oxford University Press, 2010.

\bibitem[Ghirardi et~al.(1986)Ghirardi, Rimini, and Weber]{ghirardi1986unified}
Gian~Carlo Ghirardi, Alberto Rimini, and Tullio Weber.
\newblock Unified dynamics for microscopic and macroscopic systems.
\newblock \emph{Physical Review D}, 34\penalty0 (2):\penalty0 470, 1986.

\bibitem[Diosi(1987)]{diosi1987universal}
Lajos Diosi.
\newblock A universal master equation for the gravitational violation of
  quantum mechanics.
\newblock \emph{Physics Letters A}, 120\penalty0 (8):\penalty0 377--381, 1987.

\bibitem[Ghirardi et~al.(1990)Ghirardi, Pearle, and Rimini]{ghirardi1990markov}
Gian~Carlo Ghirardi, Philip Pearle, and Alberto Rimini.
\newblock Markov processes in {H}ilbert space and continuous spontaneous
  localization of systems of identical particles.
\newblock \emph{Physical Review A}, 42\penalty0 (1):\penalty0 78, 1990.

\bibitem[Penrose(1996)]{penrose1996gravity}
Roger Penrose.
\newblock On gravity's role in quantum state reduction.
\newblock \emph{General Relativity and Gravitation}, 28\penalty0 (5):\penalty0
  581--600, 1996.

\bibitem[Chalmers and McQueen(Forthcoming, expected 2021)]{cmdraft}
D.~Chalmers and K.~McQueen.
\newblock Consciousness and the collapse of the wave function.
\newblock In S.~Gao, editor, \emph{Consciousness and Quantum Mechanics}. Oxford
  University Press, Forthcoming, expected 2021.

\bibitem[Pearle(book in preparation)]{pearle2019dynamical}
Philip Pearle.
\newblock Dynamical collapse.
\newblock In Philip Pearle, Simon Saunders, and Antony Valentini, editors,
  \emph{Quantum Theory and Reality: Pilot Waves, Dynamical Collapse, Many
  Worlds}. Oxford University Press, book in preparation.

\bibitem[Oizumi et~al.(2014)Oizumi, Albantakis, and
  Tononi]{oizumi2014phenomenology}
Masafumi Oizumi, Larissa Albantakis, and Giulio Tononi.
\newblock From the phenomenology to the mechanisms of consciousness: integrated
  information theory 3.0.
\newblock \emph{PLoS Comput Biol}, 10\penalty0 (5):\penalty0 e1003588, 2014.

\bibitem[Kent(2012)]{kent2012quantum}
Adrian Kent.
\newblock Quantum tasks in minkowski space.
\newblock \emph{Classical and Quantum Gravity}, 29\penalty0 (22):\penalty0
  224013, 2012.

\bibitem[Kent(2005{\natexlab{a}})]{kent2005nonlinearity}
Adrian Kent.
\newblock Nonlinearity without superluminality.
\newblock \emph{Physical Review A}, 72\penalty0 (1):\penalty0 012108,
  2005{\natexlab{a}}.

\bibitem[Kent(2005{\natexlab{b}})]{kent2005causal}
Adrian Kent.
\newblock Causal quantum theory and the collapse locality loophole.
\newblock \emph{Physical Review A}, 72\penalty0 (1):\penalty0 012107,
  2005{\natexlab{b}}.

\bibitem[Popescu and Rohrlich(1994)]{popescu1994quantum}
Sandu Popescu and Daniel Rohrlich.
\newblock Quantum nonlocality as an axiom.
\newblock \emph{Foundations of Physics}, 24\penalty0 (3):\penalty0 379--385,
  1994.

\bibitem[Van~Dam(2013)]{van2013implausible}
Wim Van~Dam.
\newblock Implausible consequences of superstrong nonlocality.
\newblock \emph{Natural Computing}, 12\penalty0 (1):\penalty0 9--12, 2013.

\bibitem[Brassard et~al.(2006)Brassard, Buhrman, Linden, M{\'e}thot, Tapp, and
  Unger]{brassard2006limit}
Gilles Brassard, Harry Buhrman, Noah Linden, Andr{\'e}~Allan M{\'e}thot, Alain
  Tapp, and Falk Unger.
\newblock Limit on nonlocality in any world in which communication complexity
  is not trivial.
\newblock \emph{Physical Review Letters}, 96\penalty0 (25):\penalty0 250401,
  2006.

\bibitem[Popescu(2014)]{popescu2014nonlocality}
Sandu Popescu.
\newblock Nonlocality beyond quantum mechanics.
\newblock \emph{Nature Physics}, 10\penalty0 (4):\penalty0 264--270, 2014.

\bibitem[Abrams and Lloyd(1998)]{abrams1998nonlinear}
Daniel~S Abrams and Seth Lloyd.
\newblock Nonlinear quantum mechanics implies polynomial-time solution for
  {NP}-complete and\#{P} problems.
\newblock \emph{Physical Review Letters}, 81\penalty0 (18):\penalty0 3992,
  1998.

\bibitem[Elitzur and Vaidman(1993)]{elitzur1993quantum}
Avshalom~C Elitzur and Lev Vaidman.
\newblock Quantum mechanical interaction-free measurements.
\newblock \emph{Foundations of Physics}, 23\penalty0 (7):\penalty0 987--997,
  1993.

\bibitem[White et~al.(1998)White, Mitchell, Nairz, and
  Kwiat]{white1998interaction}
Andrew~G White, Jay~R Mitchell, Olaf Nairz, and Paul~G Kwiat.
\newblock ``{I}nteraction-free'' imaging.
\newblock \emph{Physical Review A}, 58\penalty0 (1):\penalty0 605, 1998.

\bibitem[Chalmers and McQueen()]{cmtalks}
D.~Chalmers and K.~McQueen.
\newblock Consciousness and the collapse of the wave function: Presentations.
\newblock URL \url{http://consc.net/qm/}.

\bibitem[Kremnizer and Ranchin(2015)]{kremnizer2015integrated}
Kobi Kremnizer and Andr{\'e} Ranchin.
\newblock Integrated information-induced quantum collapse.
\newblock \emph{Foundations of Physics}, 45\penalty0 (8):\penalty0 889--899,
  2015.

\bibitem[Ok{\'o}n and Sebasti{\'a}n(2016)]{okon2016back}
Elias Ok{\'o}n and Miguel~Angel Sebasti{\'a}n.
\newblock How to back up or refute quantum theories of consciousness.
\newblock \emph{Mind and Matter}, 14\penalty0 (1):\penalty0 25--49, 2016.

\bibitem[Moller(1963)]{moller1963theories}
C~Moller.
\newblock Les theories relativistes de la gravitation colloques internationaux
  {CNRX} 91 edited by {A L}ichnerowicz and {M.-A. T}onnelat ({P}aris:
  {CNRS})(1962).
\newblock \emph{Nucl. Phys}, 40:\penalty0 353, 1963.

\bibitem[Rosenfeld(1963)]{rosenfeld1963quantization}
Leon Rosenfeld.
\newblock On quantization of fields.
\newblock \emph{Nuclear Physics}, 40:\penalty0 353--356, 1963.

\bibitem[Kibble(1978)]{kibble1978relativistic}
TWB Kibble.
\newblock Relativistic models of nonlinear quantum mechanics.
\newblock \emph{Communications in Mathematical Physics}, 64\penalty0
  (1):\penalty0 73--82, 1978.

\bibitem[Kibble and Randjbar-Daemi(1980)]{kibble1980non}
TWB Kibble and S~Randjbar-Daemi.
\newblock Non-linear coupling of quantum theory and classical gravity.
\newblock \emph{Journal of Physics A: Mathematical and General}, 13\penalty0
  (1):\penalty0 141, 1980.

\bibitem[Carney et~al.(2019)Carney, Stamp, and Taylor]{carney2019tabletop}
Daniel Carney, Philip~CE Stamp, and Jacob~M Taylor.
\newblock Tabletop experiments for quantum gravity: a user's manual.
\newblock \emph{Classical and Quantum Gravity}, 36\penalty0 (3):\penalty0
  034001, 2019.

\bibitem[Tilloy and Di{\'o}si(2016)]{tilloy2016sourcing}
Antoine Tilloy and Lajos Di{\'o}si.
\newblock Sourcing semiclassical gravity from spontaneously localized quantum
  matter.
\newblock \emph{Physical Review D}, 93\penalty0 (2):\penalty0 024026, 2016.

\bibitem[Kent(2020)]{kentgravexpt}
Adrian Kent.
\newblock Tests of quantum gravity near measurement events.
\newblock \emph{arxiv:2010.11811}, 2020.

\bibitem[Kent(2018)]{kent2018simple}
Adrian Kent.
\newblock Simple refutation of the {E}ppley-{H}annah argument.
\newblock \emph{Classical and Quantum Gravity}, 35\penalty0 (24):\penalty0
  245008, 2018.

\bibitem[Gisin(1990)]{gisin1990weinberg}
Nicolas Gisin.
\newblock Weinberg's non-linear quantum mechanics and supraluminal
  communications.
\newblock \emph{Physics Letters A}, 143\penalty0 (1-2):\penalty0 1--2, 1990.

\bibitem[Page and Geilker(1981)]{page1981indirect}
Don~N Page and CD~Geilker.
\newblock Indirect evidence for quantum gravity.
\newblock \emph{Physical Review Letters}, 47\penalty0 (14):\penalty0 979, 1981.

\bibitem[Ralph et~al.(2009)Ralph, Milburn, and Downes]{ralph2009quantum}
Timothy~C Ralph, Gerard~J Milburn, and T~Downes.
\newblock Quantum connectivity of space-time and gravitationally induced
  decorrelation of entanglement.
\newblock \emph{Physical Review A}, 79\penalty0 (2):\penalty0 022121, 2009.

\bibitem[Ralph and Pienaar(2014)]{ralph2014entanglement}
TC~Ralph and J~Pienaar.
\newblock Entanglement decoherence in a gravitational well according to the
  event formalism.
\newblock \emph{New Journal of Physics}, 16\penalty0 (8):\penalty0 085008,
  2014.

\bibitem[Pienaar et~al.(2011)Pienaar, Myers, and Ralph]{pienaar2011quantum}
JL~Pienaar, CR~Myers, and TC~Ralph.
\newblock Quantum fields on closed timelike curves.
\newblock \emph{Physical Review A}, 84\penalty0 (6):\penalty0 062316, 2011.

\bibitem[Joshi et~al.(2018)Joshi, Pienaar, Ralph, Cacciapuoti, McCutcheon,
  Rarity, Giggenbach, Lim, Makarov, Fuentes, et~al.]{joshi2018space}
Siddarth~Koduru Joshi, Jacques Pienaar, Timothy~C Ralph, Luigi Cacciapuoti,
  Will McCutcheon, John Rarity, Dirk Giggenbach, Jin~Gyu Lim, Vadim Makarov,
  Ivette Fuentes, et~al.
\newblock Space {QUEST} mission proposal: experimentally testing decoherence
  due to gravity.
\newblock \emph{New Journal of Physics}, 20\penalty0 (6):\penalty0 063016,
  2018.

\bibitem[Xu et~al.(2019)Xu, Ma, Ren, Yong, Ralph, Liao, Yin, Liu, Cai, Han,
  et~al.]{xu2019satellite}
Ping Xu, Yiqiu Ma, Ji-Gang Ren, Hai-Lin Yong, Timothy~C Ralph, Sheng-Kai Liao,
  Juan Yin, Wei-Yue Liu, Wen-Qi Cai, Xuan Han, et~al.
\newblock Satellite testing of a gravitationally induced quantum decoherence
  model.
\newblock \emph{Science}, 366\penalty0 (6461):\penalty0 132--135, 2019.

\bibitem[Bose et~al.(2017)Bose, Mazumdar, Morley, Ulbricht, Toro{\v{s}},
  Paternostro, Geraci, Barker, Kim, and Milburn]{bose2017spin}
Sougato Bose, Anupam Mazumdar, Gavin~W Morley, Hendrik Ulbricht, Marko
  Toro{\v{s}}, Mauro Paternostro, Andrew~A Geraci, Peter~F Barker, MS~Kim, and
  Gerard Milburn.
\newblock Spin entanglement witness for quantum gravity.
\newblock \emph{Physical Review Letters}, 119\penalty0 (24):\penalty0 240401,
  2017.

\bibitem[Marletto and Vedral(2017)]{marletto2017gravitationally}
Chiara Marletto and Vlatko Vedral.
\newblock Gravitationally induced entanglement between two massive particles is
  sufficient evidence of quantum effects in gravity.
\newblock \emph{Physical Review Letters}, 119\penalty0 (24):\penalty0 240402,
  2017.

\bibitem[Marshman et~al.(2020)Marshman, Mazumdar, and
  Bose]{marshman2020locality}
Ryan~J Marshman, Anupam Mazumdar, and Sougato Bose.
\newblock Locality and entanglement in table-top testing of the quantum nature
  of linearized gravity.
\newblock \emph{Physical Review A}, 101\penalty0 (5):\penalty0 052110, 2020.

\end{thebibliography}
\end{document}